\documentclass[10pt,journal]{IEEEtran}
\usepackage{amsmath,amsfonts}
\usepackage{algorithmic}
\usepackage{algorithm}
\usepackage{array}
\usepackage[caption=false,font=normalsize,labelfont=sf,textfont=sf]{subfig}
\usepackage{textcomp}
\usepackage{stfloats}
\usepackage{url}
\usepackage{color}
\usepackage{verbatim}
\usepackage{graphicx}
\usepackage{cite}
\usepackage{makecell}
\usepackage{booktabs,multirow}
\usepackage[hidelinks]{hyperref} 
\usepackage{orcidlink}
\newcommand{\mfmod}[1]{{\color{red}{\emph{#1}}}}
\renewcommand{\mfmod}[1]{} 

\usepackage{soul}

\begin{document}

\title{UniPET-SPK: A Unified Framework for Parameter-Efficient Tuning of Pre-trained Speech Models 
for Robust Speaker Verification
}

\author{Mufan Sang\textsuperscript{\large\orcidlink{0000-0003-2510-3148}},~\IEEEmembership{Student Member,~IEEE}, and John H. L. Hansen\textsuperscript{\large\orcidlink{0000-0003-1382-9929}},~\IEEEmembership{Fellow,~IEEE}}

\markboth{Journal of \LaTeX\ Class Files,~Vol.~14, No.~8, August~2021}%
{Shell \MakeLowercase{\textit{et al.}}: A Sample Article Using IEEEtran.cls for IEEE Journals}

\maketitle

\begin{abstract}
With excellent generalization ability, self-supervised speech models have shown impressive performance on various downstream speech tasks in the pre-training and fine-tuning paradigm. However, as the size of pre-trained models grows, fine-tuning becomes practically unfeasible due to expanding computation and storage requirements, as well as the risk of overfitting. In this study, we concentrate on exploring parameter-efficient tuning (PET) methods for adapting large-scale pre-trained self-supervised speech models to the speaker verification task. Correspondingly, we propose three parameter-efficient tuning methods: (i) an adapter-tuning method, (ii) a prompt-tuning method, and (iii) a unified framework that effectively incorporates adapter-tuning and prompt-tuning with a dynamically learnable gating mechanism. First, we propose the Inner+Inter Adapter framework, which inserts two types of adapters into pre-trained models, allowing for adaptation of latent features within the intermediate Transformer layers and output embeddings from all Transformer layers, through a parallel adapter design. Second, we propose the Deep Speaker Prompting method that concatenates trainable prompt tokens into the input space of pre-trained models to guide adaptation. Lastly, we propose the UniPET-SPK, a unified framework that effectively incorporates these two alternate PET methods into a single framework with a dynamic trainable gating mechanism. The proposed UniPET-SPK learns to find the optimal mixture of PET methods to match different datasets and scenarios. We conduct a comprehensive set of experiments on several datasets to validate the effectiveness of the proposed parameter-efficient tuning methods. Experimental results on the VoxCeleb, CN-Celeb, and the 1$^{\rm{st}}$48-UTD forensic datasets demonstrate that the proposed UniPET-SPK can consistently outperform the two PET methods, fine-tuning, and other parameter-efficient tuning methods, achieving superior performance while updating only 5.4\% of the parameters. We further conduct experiments on the CN-Celeb and 1$^{\rm{st}}$48-UTD datasets to demonstrate the robustness and generalization ability of the proposed methods for speaker verification in different languages and more challenging scenarios.  

\end{abstract}

\begin{IEEEkeywords}
Speaker verification, pre-trained model, adapter-tuning, prompt-tuning, gating mechanism, transfer learning, parameter-efficient tuning
\end{IEEEkeywords}

\section{Introduction}
\IEEEPARstart{A}{UTOMATIC} Speaker Verification (ASV) is a task aimed at identifying the true characteristics of a speaker and accepting or discarding the identity claimed by the speaker. In recent years, extensive advancements in speaker verification have been driven by deep learning. Recent advancements in ASV are mainly attributed to diverse models and methods from supervised ASV systems encompassing different DNN architectures~\cite{snyder2018x, zeinali2019but, desplanques2020ecapa}, attention mechanisms~\cite{zhou2019deep, yadav2020frequency, sang22_interspeech, qin2022simple}, Transformer-based architectures~\cite{mary2021s, zhang22h_interspeech, sang2023improving}, to self-supervised ASV systems~\cite{chen2022unispeech, zhang2021contrastive, sang2022self}. Domain mismatch is also one of the challenging problems in this field~\cite{bhattacharya2019generative, wang20w_interspeech, sang2021deaan}. As one application of speaker verification, forensic-related problems are more complex with unique challenges due to naturalistic field recordings, diversity domain mismatch, location uncertainty, constrained availability of speech data, and short duration utterances~\cite{hansen2015speaker, poddar2018speaker, sang2020open, wang2021multi}. Most of these studies focus on utilizing task-specific datasets to train small-scale ASV systems from scratch. Recently, the emergence of large-scale pre-trained speech models has propelled research in the field of speech processing. Taking advantage of the Transformer architecture~\cite{NIPS2017_3f5ee243}, self-supervised learning (SSL), and increasingly large amounts of unlabeled data, pre-trained models have exhibited strong generalization capabilities and re-usability across various downstream speech tasks. Applying large-scale pre-trained speech models (e.g., wav2vec 2.0~\cite{baevski2020wav2vec}, HuBERT~\cite{hsu2021hubert}, and WavLM~\cite{chen2022wavlm}) to downstream tasks achieves competitive performance and even outperforms conventional task-specific models. The question of how to more efficiently utilize pre-trained models to improve the performance of downstream tasks remains an open area for investigation. 

\begin{figure}[t]
\flushright
\scalebox{1.05}
{
\hspace{-5mm} \includegraphics[width=9.1cm,height=6.7cm]{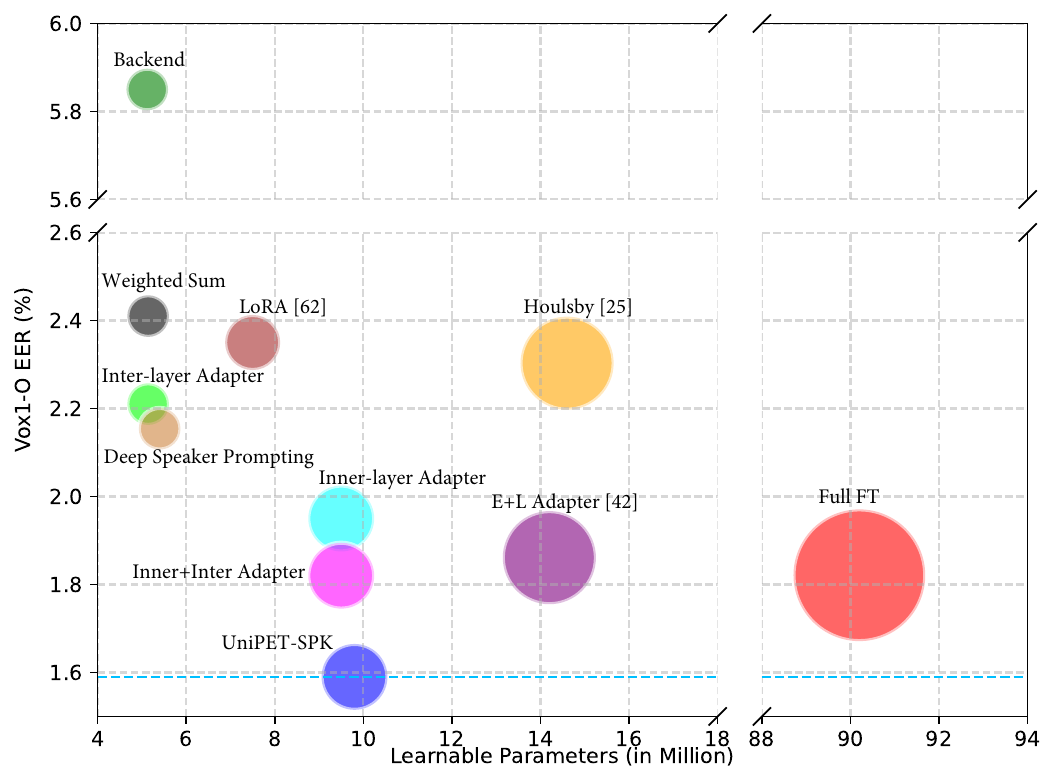}
}
\caption{Performance of parameter-efficient tuning approaches on Vox1-O. The area of each circle is proportional to the total number of tunable parameters including SV backend.} 
\label{fig:circle}
\end{figure} 

The pre-training and fine-tuning paradigm has emerged as the predominant strategy for adapting pre-trained models to downstream speech tasks. However, fine-tuning presents significant challenges due to two main reasons. First, fine-tuning requires that we update all model parameters as well as store and deploy a separate copy of the model parameters for each individual downstream task. With the increasing size of pre-trained SSL models, fine-tuning becomes prohibitively costly in terms of training, storage, and deployment, rendering it practically infeasible. Second, large-scale pre-trained models are prone to overfitting when fine-tuned with limited data for specific downstream tasks, which in turn also degrades their generalization capabilities. Therefore, exploring parameter-efficient fine-tuning methods is crucial for large-scale pre-trained model adaptation when new domains or audio capture and environments are encountered. In addition to fine-tuning, a simple and straightforward approach is linear probing, which involves keeping the pre-trained model fixed while only updating the stacked task-specific classification head or backend for each downstream task. However, this method often results in unsatisfactory performance compared to full fine-tuning.

\begin{figure*}[t]
\centering
\scalebox{1.0}
{
\includegraphics[width=15.2cm,height=7.3cm]{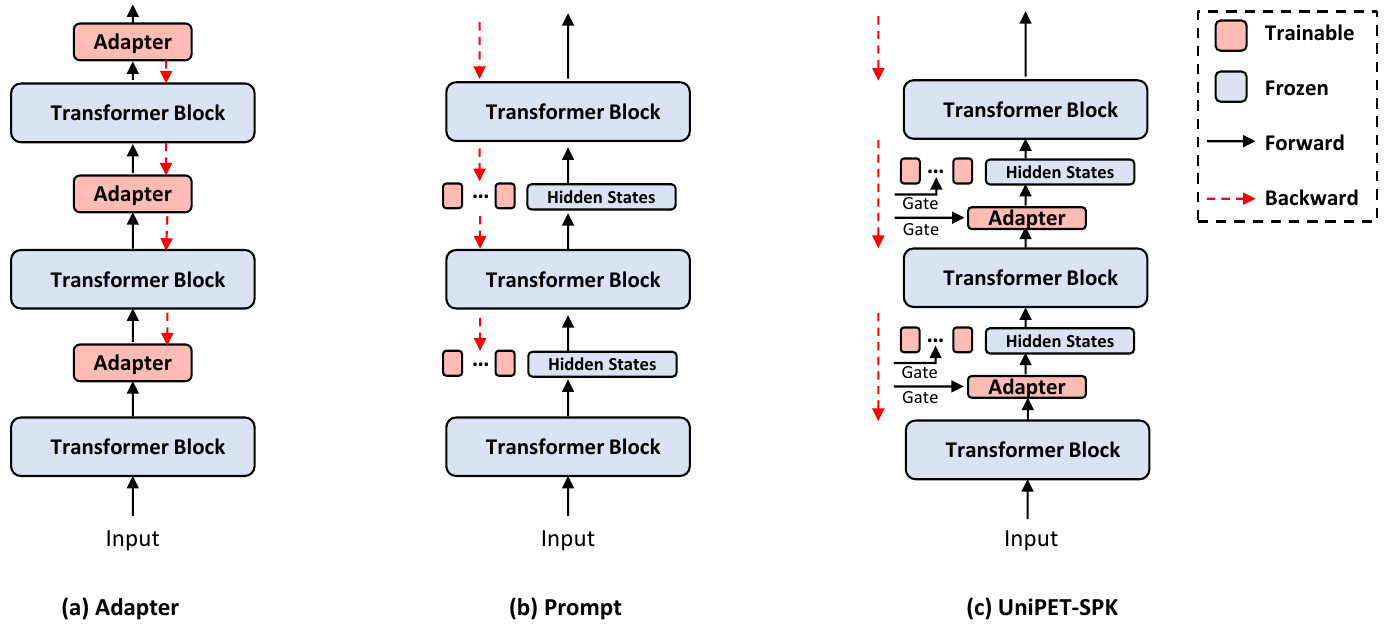}
}
\vspace{-3.0mm}
\caption{Overview of the parameter-efficient tuning methods. (a) Adapter, which inserts learnable lightweight networks into a pre-trained model; (b) Prompt, which adds learnable continuous vectors to warp the hidden states; and (c) UniPET-SPK, which dynamically incorporates Adapter and Prompt with a dynamic gating mechanism.} 
\label{fig:overview fig}
\end{figure*} 

Recently, adapters have increasingly drawn attention for transferring knowledge of pre-trained models to downstream tasks. Adapter~\cite{houlsby2019parameter} was initially introduced in the field of Natural Language Processing (NLP) for adapting pre-trained language models (PLMs). Adapters involve inserting lightweight modules with a bottleneck architecture into Transformer layers after both multi-head self-attention (MHSA) and feed-forward network (FFN) modules. These modules are typically characterized by a pair of down and up projections that shrink and recover the size of the hidden states. During fine-tuning, only the adapters get updated, leaving the rest of the model fixed. Some recent studies~\cite{thomas2022efficient,fan2022draft,chen2023chapter} have explored the use of adapters to adapt pre-trained models to diverse speech processing tasks. Nevertheless, most fail to adequately leverage the information embedded in different layers within the pre-trained models. In our previous work~\cite{10446686}, we proposed an adapter-tuning framework and explored an efficient way to utilize speaker-related information. More efficient methods for adapting pre-trained models to speaker verification are worthwhile to explore. 

Meanwhile, prompting-based methods have been proposed in the NLP field for transfer learning of large language models (LLMs), as a parallel branch of PET methods~\cite{PTSurvey}. Prompt-based learning involves designing textual inputs for models with either templated or learnable prompts that incorporate task-specific information. Among them, prefix-tuning and prompt-tuning~\cite{li2021prefix, lester2021power} add continuous prompts which are task-specific learnable vectors to the input space or the latent space of pre-trained models. When fine-tuning is performed on specific downstream datasets, only these added prompts are updated, while the entirety of the pre-trained model remains frozen.

However, adapter-based tuning and prompt-based tuning exhibit distinct mechanisms when adapting pre-trained models. While adapters-based methods inject neural modules into Transformer layers, prompt-based methods actually warp the input of Transformer layers with additional context or learnable continuous vectors. Regarding large-scale pre-trained models, different encoder layers embed different information related to various downstream tasks, and different PET methods cannot always perform well on different tasks~\cite{mao2022unipelt,peng2023parameter}. It is nontrivial to select the most appropriate PET method for a specific downstream task. Furthermore, a question naturally arises: is it necessary for each layer of a pre-trained model to employ the same PET method to achieve optimal performance? To answer this question, we explore a unified PET framework that enables flexible assignment of a mixture of PET methods to each layer rather than relying on a fixed strategy.

In this study, we explore parameter-efficient fine-tuning methods for speaker verification in general scenarios and more challenging forensic scenarios. As shown in Fig. \ref{fig:overview fig}, we propose three approaches: an adapter-based method, a prompt-based method, and a unified parameter-efficient tuning framework. We first introduce the effective adapter-tuning method which consists of two modules: (i) the Inner-layer Adapter and the (ii) Inter-layer Adapter. The former is designed to be inserted within Transformer layers for adapting latent representations within intermediate Transformer layers. The latter is added after pre-trained models to adapt the merged output embeddings from all Transformer layers. The aim is to efficiently transfer the universal knowledge of pre-trained SSL model to the speaker verification task. \mfmod{The proposed adapters learn task-specific knowledge for speaker verification by adapting latent representations within intermediate Transformer layers and the merged output embeddings from all Transformer layers. }Additionally, we present a parallel adapter design with a scaling operation to control the adapter outputs and balance task-agnostic and task-specific features learned from the original frozen FFN branch and the adapter branch within each Transformer block. Secondly, we propose Deep Speaker Prompting, which prepends a small set of trainable vectors to the input space of each Transformer layer, thereby guiding pre-trained models in learning to extract accurate speaker embeddings. Lastly, we propose UniPET-SPK, a unified parameter-efficient tuning framework for speaker verification, which incorporates our adapters and Deep Speaker Prompting dynamically through a learnable gating mechanism. The gating mechanism learns to amplify the impact of the submodule which contributes more to the current task or dataset. In this way, UniPET-SPK learns to find the most appropriate mixture setting of the two methods for each Transformer layer of pre-trained models. As shown in Fig. \ref{fig:circle}, experimental results demonstrate that the proposed UniPET-SPK surpasses our adapter-tuning method, prompt-tuning method, full fine-tuning, and other transfer learning methods while updating only 5.4\% of the parameters. 

In summary, the main contributions of this study are summarized as follows: 
\begin{itemize}
\item{We demonstrate that when adapting a large-scale pre-trained speech model on speaker verification, PET methods can achieve competitive, and even better performance, than full fine-tuning with much fewer parameters. Dynamically incorporating alternative PET methods can bring benefits and further improve performance, as shown in Fig. \ref{fig:circle}.}

\item{We introduce an efficient adapter-tuning method with a parallel architecture and the Deep Speaker Prompting method to efficiently transfer knowledge of well-generalized pre-trained speech models to the speaker verification task.}
\item{We propose UniPET-SPK, a unified parameter-efficient tuning framework that dynamically incorporates the above two methods with a trainable gating mechanism, and learn to find the most appropriate mixture of these two methods for every encoder layer for our speaker verification task.}
\item{We conduct extensive experiments on VoxCeleb datasets to investigate the effectiveness and efficiency of our proposed methods. To investigate the robustness and generalization ability of proposed methods, we further conduct experiments in the cross-lingual scenario with CN-Celeb dataset and the more complex naturalistic forensic speaker verification scenario with the $1^{\rm{st}}$48-UTD corpus. The UniPET-SPK is shown to consistently outperform full fine-tuning and other PET methods while only updating 5.4\% of the parameters.}  
 
\end{itemize}

\section{RELATED WORK}
\subsection{Self-supervised Pre-trained Speech Models}

Self-supervised learning (SSL) has gained much attention in speech research for its ability to leverage vast amounts of unlabeled data, facilitating the learning of generic representations. Recently, various SSL based pre-trained models and methods, including wav2vec~\cite{schneider2019wav2vec}, wav2vec 2.0~\cite{baevski2020wav2vec}, HuBERT~\cite{hsu2021hubert}, and WavLM~\cite{chen2022wavlm} have been shown to be effective on various speech tasks. Among them, WavLM was proposed to explore full-stack speech tasks instead of focusing on specific tasks. WavLM combines masked speech prediction and denoising in pre-training to learn not only automatic speech recognition (ASR) knowledge but also information to address additional non-ASR tasks. In summary, these SSL pre-trained models have advanced the development of various speech fields with excellent generalization ability. The pre-training and fine-tuning paradigm has further facilitated the adaptation of pre-trained models to downstream speech tasks. However, fine-tuning large-scale pre-trained models is still data-dependent and computationally expensive, which limits wider application of SSL pre-trained models. Therefore, it is worthwhile to explore how to more efficiently adapt pre-trained models to downstream tasks with lower computation and storage costs. 

\subsection{Adapter-based Tuning}
Adapters~\cite{houlsby2019parameter} have been introduced as an alternative approach for adapting large-scale pre-trained language models in NLP. Adapters modify the feature extractors by inserting one or more lightweight bottleneck modules without changing the parameters of the pre-trained models. Adapter-based methods have been shown to be comparable to ﬁne-tuning with much higher parameter efficiency, and can at times perform slightly better in low-resource scenarios~\cite{he2022towards}. With the advantage, adapters have also been applied to computer vision tasks~\cite{sung2022vl,chen2022adaptformer}  
 
Recently, adapters have also been introduced for speech processing tasks. In~\cite{kannan2019large}, researchers applied adapters to the RNN-T model for multi-lingual ASR. Also, ~\cite{winata2020adapt} proposed to apply adapters to a speech Transformer to mitigate the long-tail problem of multilingual ASR. In~\cite{thomas2022efficient}, adapters were inserted into wav2vec 2.0 to increase the model scalability to multiple languages. In~\cite{fan2022draft}, adapters were also utilized to improve domain adaptation of SSL models including wav2vec 2.0 and HuBERT for children's ASR. Moreover, the SimAdapter~\cite{hou2021exploiting} was proposed for cross-lingual low-resource ASR. Further research~\cite{chen2023chapter,otake2023parameter,peng2023parameter} also investigated the effectiveness of adapters for different downstream speech tasks beyond ASR (e.g., emotion recognition, speaker verification, intent classiﬁcation). However, most of these studies do not sufficiently utilize the task-related information embedded in the hidden layers of the pre-trained models. In our study, we first introduce an efficient and effective adapter framework for speaker verification.

\subsection{Prompt-Tuning}
Prompting has been proposed in the field of NLP for GPT-series models~\cite{radford2019language, NEURIPS2020_1457c0d6}. This technique involves prepending specific language instructions or task-relevant descriptions to the input text, enabling a pre-trained large-scale language model to understand and generalize across downstream tasks. However, there are limitations that rely on hand-crafted design, as well as the need to design prompting requirements manually for different specific tasks. Hand-crafted prompts are also difficult to optimize and are inherently limited in the number of training examples by the maximum model input length. To overcome these limitations, recent studies have explored the use of soft prompts~\cite{liu2023gpt,lester2021power,li2021prefix, liu2022p, jia2022visual}, where continuous prompts are prepended to the model’s input embeddings or layers and optimized via gradient descent while keeping the whole backbone frozen. This shifts the challenge from creating discrete prompts to continuous optimization. Compared to full fine-tuning, prompt-tuning achieves competitive performance however only with updates of far fewer parameters. 

Although prompt tuning has been applied to language and vision models~\cite{pmlr-v139-radford21a,zhou2022learning}, its potential in the speech processing domain remains underexplored and worthwhile to explore. Chang et al.~\cite{chang22e_interspeech} proposed to apply prompt tuning on Generative Spoken Language Model to perform speech classification and sequence generation tasks. To further investigate the generalizability of the prompting paradigm, SpeechPrompt v2~\cite{chang2023speechprompt} was proposed for several speech classification tasks such as audio classification, speech command recognition, emotion recognition, and others. Concurrently, researchers~\cite{kim2022integrated} explored the efficient transfer learning method on Audio Spectrogram Transformer (AST)~\cite{gong21b_interspeech} and Wav2vec 2.0 model for several downstream tasks. Also, other studies~\cite{peng2023parameter,LiDual} analyzed parameter-efficient transfer learning methods to pre-trained Transformer models for speaker verification. Despite these advancements, research on how to design a unified framework of PET for integrating adapter-tuning and prompt-tuning dynamically is lacking in the speech field. Most of the previous research primarily employ PET methods independently. In this study, we explore a unified parameter-efficient transfer learning framework that incorporates different PET methods as submodules and learns to dynamically adjust the mixture of PET methods to better align with the current task-specific dataset or domain.

\begin{figure*}[th]
\centering
\scalebox{1.00}
{
\includegraphics[width=15.5cm,height=10.0cm]{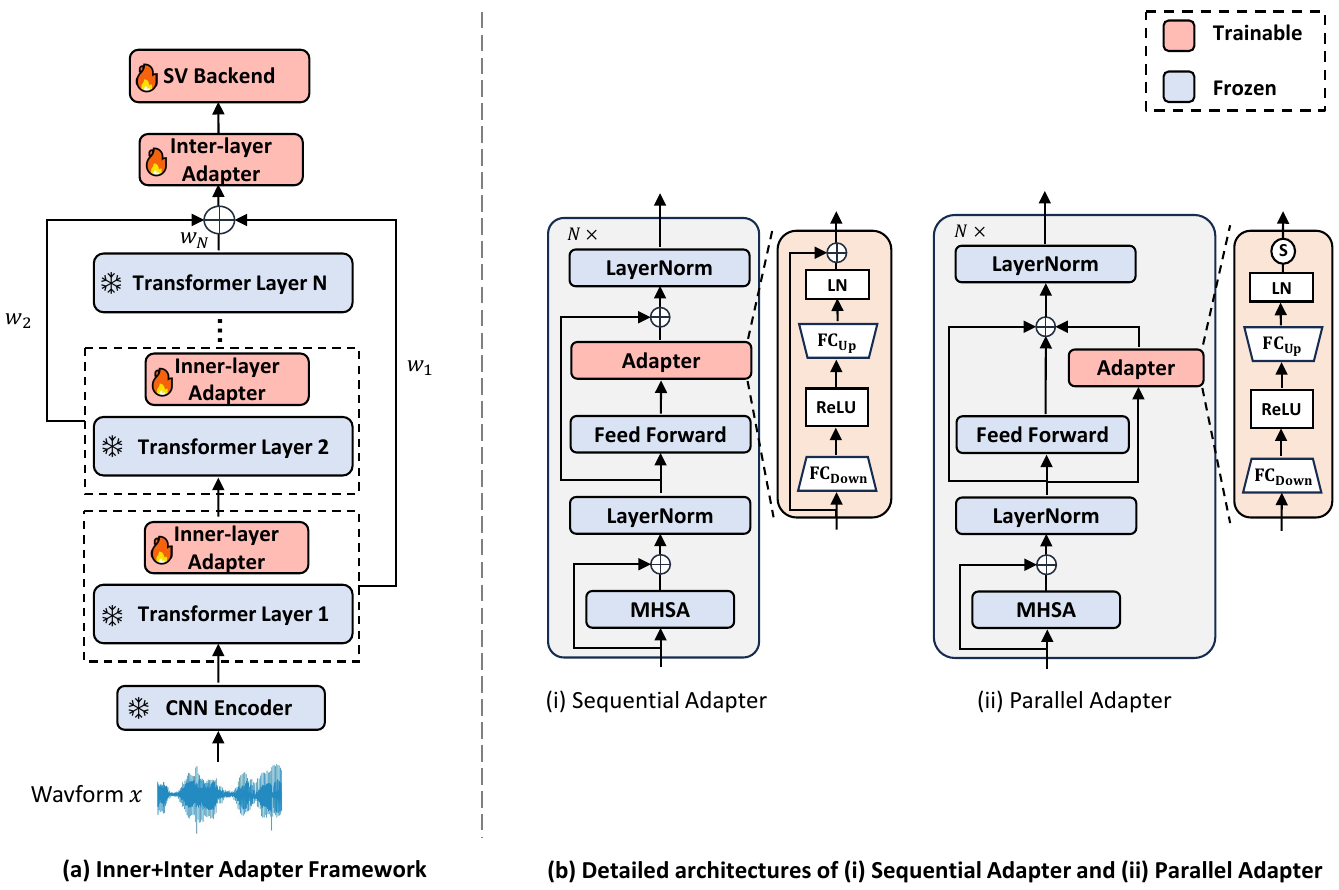}
}
\caption{Overview of the pre-trained model with Inner+Inter Adapter framework, and the detailed architectures of sequential and parallel adapters. During fine-tuning, the pre-trained model is frozen, only the Inner-layer Adapter, Inter-layer Adapter, and the SV backend are updated.} 
\label{fig:system}
\end{figure*}

\vspace{1.0ex}
\section{SPEAKER ADAPTER TUNING}
In this section, we introduce our adapter-tuning framework designed to efﬁciently transfer knowledge of large pre-trained speech models to the task of speaker verification. This approach integrates two types of adapters into a frozen pre-trained backbone: (1) the Inner-layer Adapter, inserted within the intermediate Transformer layers, and (2) the Inter-layer Adapter, inserted after the weighed-sum operation between the pre-trained backbone and speaker verification backend. \mfmod{inserted the inner-layer adapters are inserted within the intermediate Transformer layers and the inter-layer adapter is inserted after the weighted-sum operation between the pre-trained backbone and speaker verification backend. }The overall system framework is shown in Fig. \ref{fig:system}(a).

\subsection{Inner-layer Adapter and Inter-layer Adapter}
 
\textbf{Inner-layer Adapter}: Adapters are lightweight modules injected into Transformer layers of pre-trained models for transfer learning. To keep the generalization ability of pre-trained models, adapter-tuning methods typically fine-tune the adapters only while keeping pre-trained models frozen during training. A prior study~\cite{yang2021superb} indicated that leveraging output representations from the lower layers of the pre-trained models contributes to better performance for downstream speech tasks. To better utilize speaker-related information embedded in all intermediate layers, we propose the Inner-layer Adapter, facilitating the adaptation of latent features within the intermediate Transformer layers.

Different from previous studies~\cite{houlsby2019parameter,thomas2022efficient}, which employed adapters following both the MHSA and FFN modules, we insert the Inner-layer Adapter exclusively after the FFN to enhance parameter efficiency. The Inner-layer Adapter has a bottleneck structure comprising a down projection from hidden dimension $d$ to bottleneck dimension $\hat{d}$ with parameter $\boldsymbol{W}_{\text {down}}\in R^{d \times \hat{d}}$, with an up project represented by parameter $\boldsymbol{W}_{\text {up}}\in R^{\hat{d} \times d}$, a non-linear activation function situated between them, and a residual connection. For an input feature $\boldsymbol{x_{i}}$ of the FFN, the output of the Inner-layer Adapter is formulated as:

\begin{equation}
\begin{aligned}
\tilde{\boldsymbol{z}^s_i} &=\mathrm{FFN}\left(\boldsymbol{x_i}\right)+\mathrm{LN}( \boldsymbol{W}_{\text {up}}f\left(\boldsymbol{W}_{\text {down }}\mathrm{FFN}\left(\boldsymbol{x_i}\right)\right)),
\end{aligned}
\end{equation}
where $f$ denotes the ReLU activation function. 


\textbf{Inter-layer Adapter}: Similar to an earlier study~\cite{yang2021superb}, we add a group of learnable weights to average the hidden representations across all encoder layers. Inner-layer Adapters are integrated into the intermediate encoder layers in order to adapt the latent features within layers explicitly. However, we note that the interaction amongst the encoder layers is not directly considered. To better adapt the pre-trained model and fully leverage the speaker-related information embedded in all hidden states, we propose our Inter-layer Adapter. As shown in Fig. \ref{fig:system}(a), we insert the Inter-layer Adapter after the weighted sum operation. The Inter-layer Adapter consists of a fully connected (FC) layer, followed by a non-linear activation function and layer normalization (LN). Given the output representation from the $i$-th layer as $\boldsymbol H_{i}$, the output generated by the Inter-layer Adapter can be computed as:
\begin{equation}
\begin{aligned}
\boldsymbol{\tilde{H}}=\mathrm{LN}(f(\boldsymbol{W}_{\text {inter}}(\sum_{i=1}^{N} w_{i} \boldsymbol H_{i}\label{con:DCT1D2}))),
\end{aligned}
\end{equation}
where $\boldsymbol{W}_{\text {inter}}\in R^{d \times e}$ denotes the FC layer, and $d$ and $e$ denote the hidden dimension and speaker embedding dimension, respectively. Also, $f$ denotes the ReLU activation function where $w_{i}$ denotes the trainable weight for the $i$-th hidden state.

\subsection{Parallel Adapter Design}
Conventional adapters are usually added sequentially after MHSA and FFN modules. Inspired by studies~\cite{chen2022adaptformer,he2022towards}, we adopt a parallel adapter design which is illustrated in Fig. \ref{fig:system}(b). Contrary to the sequential adapter architecture, our parallel adapter is integrated into an additional sub-branch, focusing on the acquisition of task-specific knowledge. Furthermore, we introduce a scaling factor $s$ in order to rescale the output of the parallel adapter before adding it to the original branch via a residual connection. We aim to use the scaling factor $s$ to control the balance between the task-agnostic features obtained from the original frozen branch and the task-specific features obtained from the parallel adapter branch. This parallel design helps the pre-trained speech model preserve its generalization capability, while the adapters learn domain-specific features to serve as a complementary part for the feature ensemble. The overall output of the parallel adapter is defined as follows:
\begin{equation}
\begin{aligned}
\tilde{\boldsymbol{z}^{p}_i} &=\mathrm{LN}(\boldsymbol{W}_{\text {up}}f\left(\boldsymbol{W}_{\text {down }}\boldsymbol{x_i}\right)).
\end{aligned}
\end{equation}
Accordingly, the adapter branch, FFN branch, and the residual connection are fused before layer normalization. The final output of the $i$-th Transformer layer can then be shown as:
\begin{equation}
\begin{aligned}
{\boldsymbol H_{i}} &=\mathrm{LN}\left(\mathrm{FFN}\left(\boldsymbol{x_i}\right)+s \cdot \tilde{\boldsymbol{z}^{p}_i}+\boldsymbol{x_i}\right).
\end{aligned}
\end{equation}

\vspace{-1.0ex}
\begin{figure*}[th]
\centering
\scalebox{0.99}
{
\includegraphics[width=18.4cm,height=9.3cm]{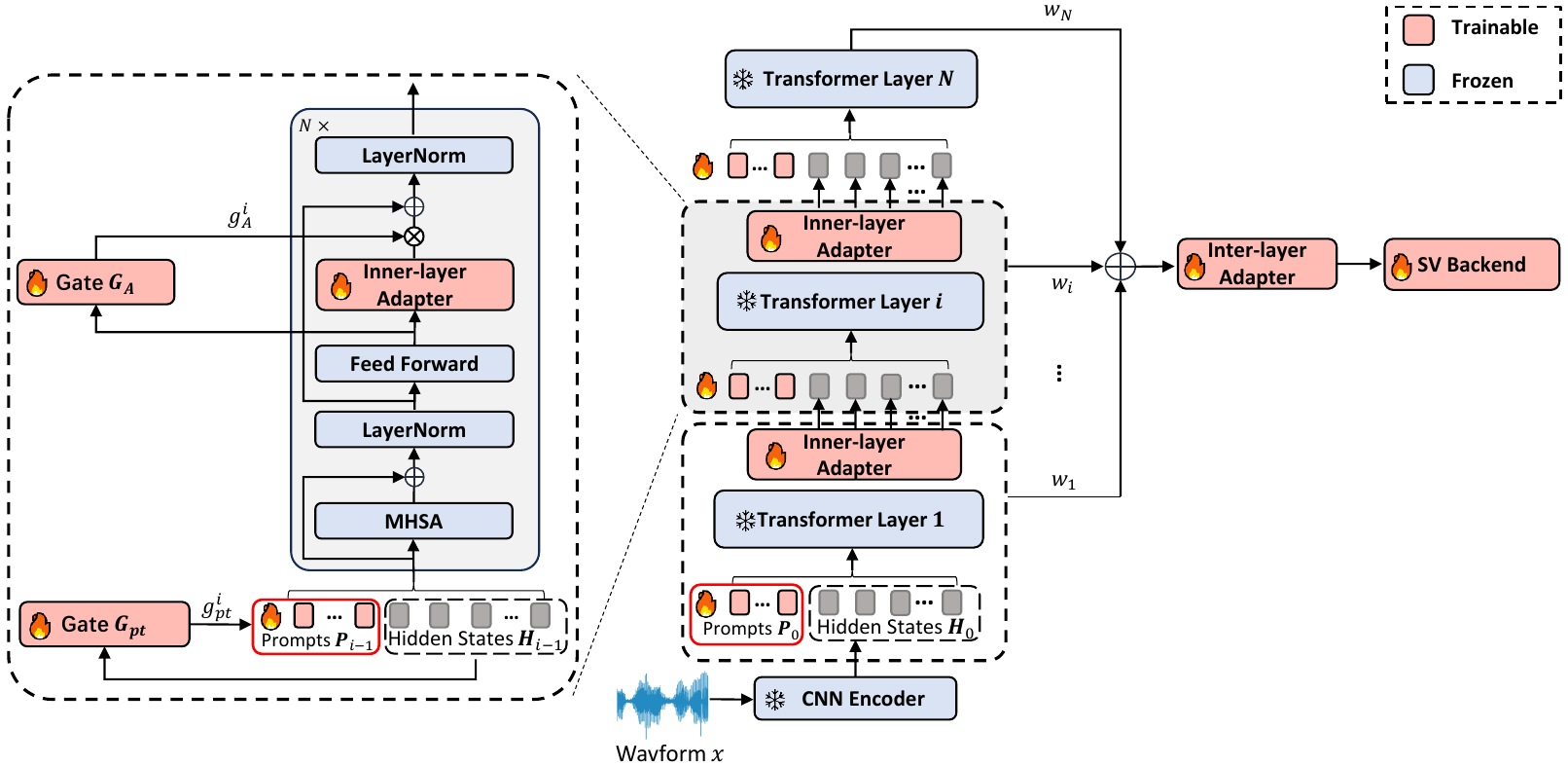}}
\caption{Overview of the UniPET-SPK framework. During fine-tuning, the pre-trained model is frozen, only the prepended speaker prompts, Inner-layer Adapter, Inter-layer Adapter, gating layers, and the SV backend are updated.} 
\label{fig:UniPET-system}
\end{figure*}

\section{DEEP SPEAKER PROMPTING}
\mfmod{Adapters typically process output latent features obtained from MHSA or FFN to make the pre-trained model be adapted to downstream tasks. The output tokens of the last MHSA blocks are unchanged, and there is no information exchange between tokens after that. Considering make changes to the token mixer of the pre-trained Transformer model, we explore a method which tune MHSA to adjust the token mixer on downstream tasks, named prompt-tuning.} 
Next, we propose Deep Speaker Prompting to adapt pre-trained speech models for speaker verification in a novel way based on task-specific prompts. Different from adapter-based methods that inject new modules into the pre-trained models, these prompt-based methods modify the original input with additional context. Prompts can be a set of learnable continuous parameters that are prepended to the input of Transformer layers, acting as task-specific instructions in order to steer the information from the fixed pre-trained model. During fine-tuning, the pre-trained model is kept frozen and only these trainable prompts are updated, with the aim to learn and capture task-specific information to instruct pre-trained model adaptation towards the downstream task. For WavLM and HuBERT, the input speech signal $x$ is first encoded by the CNN encoder and then embedded into the $d$-dimensional output representations $H_{0}$. In this method, we introduce a collection of $m$ randomly initialized task-specific prompts $\boldsymbol{P}_{0}=[p^1_0, p^2_0, ..., p^m_0]\in R^{m \times d}$, where $d$ denotes the dimension of each prompt vector. These prompts are prepended to the convolutional output representation $\boldsymbol{H}_{0}$, and serve as input for the first Transformer layer $\boldsymbol {W}^{1}_{Block}$, as formulated below: 
\begin{equation}
\begin{aligned}
\left[\boldsymbol{Z}_{1}, \boldsymbol{H}_{1}\right]&=\boldsymbol{W}^{1}_{Block}\left([\boldsymbol{P}_{0},\boldsymbol{H}_{0}]\right)\label{con:PT},
\end{aligned}
\end{equation}
\begin{equation}
\begin{aligned}
\left[\boldsymbol{Z}_{i}, \boldsymbol{H}_{i}\right]&=\boldsymbol{W}^{i}_{Block}\left([\boldsymbol{Z}_{i-1},\boldsymbol{H}_{i-1}]\right), \quad i=2,3, \ldots, N \label{con:PT2}  
\end{aligned}
\end{equation}
where $\boldsymbol{H}_{i}\in R^{T \times d}$ denotes the hidden states of the $i$-th Transformer layer corresponding to the input speech part. Here, $\boldsymbol{Z}_{i}\in R^{m \times d}$ represents the latent representation corresponding to the prompts computed by the $i$-th Transformer layer. The prompts $\boldsymbol{P}_{0}\in R^{m \times d}$ has the same dimension as the encoded feature $\boldsymbol{H}_{0}$. The hidden representation of the $i$-th layer $[\boldsymbol{Z}_{i}, \boldsymbol{H}_{i}]\in R^{(m+T) \times d}$ is the input for the $(i+1)$-th layer. Throughout the MHSA layers in the Transformer blocks, the collection of prepended prompts is able to modify the input distribution and adapt the pre-trained model to specific downstream tasks. 
 
To enhance the prompt's instruction ability, we further insert new prompt tokens into the inputs of all intermediate Transformer layers other than the first layer. We denote these learnable prompts for the $(i+1)$-th Transformer layer as $\boldsymbol{P}_{i}=[p^1_i,p^2_i,\ldots,p^m_i]\in R^{m \times d},$ $i=0,1,2,\ldots,N-1$. All new prompts are concatenated to the input of each layer and learn the task-specific instruction to achieve pre-trained model adaptation. With respect to the output hidden states of each Transformer layer, prompt feature embedding $\boldsymbol{Z}_{i}$ is discarded and new prompts are prepended to the remaining hidden states. Similar to Eq. \ref{con:PT2}, the overall formulated prompt solution is summarized here,
\begin{equation}
\begin{aligned}
\left[\boldsymbol{Z}_{i}, \boldsymbol{H}_{i}\right]&=\boldsymbol{W}^{i}_{Block}\left([\boldsymbol{P}_{i-1},\boldsymbol{H}_{i-1}]\right) \quad i=1,2, \ldots, N.
\end{aligned}
\end{equation}

In this way, we introduce new prompts at every layer to help modify the audio or speech stimulus, guiding the model towards generating representations that are tailored to specific tasks. During fine-tuning, the pre-trained model is kept frozen and only the task-specific prompts prepended at the embedding space of each layer are updated through gradient descent. Compared to full fine-tuning, Deep Speaker Prompting only requires updating the learnable continuous prompts, which is much more parameter-efficient. After fine-tuning, it suffices to retain only the task-specific prompts and the corresponding SV backend, the original copy of the pre-trained model can then be shared and re-used across different tasks or domains.

\section{A UNIFIED FRAMEWORK FOR PARAMETER-EFFICIENT TUNING IN SPEAKER VERIFICATION}
According to the proposed adapter-tuning and prompt-tuning methods, we can show that these two methods exhibit distinct processing, performance mechanisms, and characteristics. Considering that these two parameter-efficient tuning methods act on different parts of the pre-trained model and may be compatible with each other, it is possible to apply them to the same pre-trained model simultaneously. Furthermore, different Transformer layers in the pre-trained model may require other appropriate PET methods or a mixture of PET methods to achieve the best overall performance on input application domains and datasets.  

In light of this insight, we propose a unified PET framework which incorporates a dynamic gating mechanism to effectively integrate both methods to bring additional benefits and enhance system performance beyond what each method could achieve independently. As illustrated in Fig. \ref{fig:UniPET-system}, the proposed framework incorporates adapter and prompt-tuning as submodules for each Transformer layer. To determine the importance of both methods, we propose the learnable gating mechanism to control their involvement dynamically. We define the learnble gates for the Deep Speaker Prompting as $\boldsymbol G_{pt}=[g^1_{pt},g^2_{pt},...,g^N_{pt}]$, $g^i_{pt}\in(0,1)$ and the Inner+Inter Adapter as $\boldsymbol G_{A}=[g^1_{A},g^2_{A},...,g^N_{A},g^{N+1}_{A}]$, $g^i_{A}\in(0,1)$. Alternatively, the learnable gate $\boldsymbol G_{pt}$ and $\boldsymbol G_{A}$ are trained to dynamically regulate the influence of prompt tuning and adapters, respectively. Intuitively, the gates learn to increase the impact of the submodule, which is more suitable for the current task or dataset. On the other hand, this mechanism does allow our solution to be complemented with mutual benefits. 

To be more specific, for Deep Speaker Promoting, the learnable gate $g^i_{pt}$ is applied to scale prompt tokens before prepending to the input of the $i$-th Transformer layer. The gating operation for the $i$-th layer can be formulated as:
\begin{equation}
\begin{aligned}
\left[\boldsymbol{Z}_{i}, \boldsymbol{H}_{i}\right]&=\boldsymbol{W}^{i}_{Block}\left([g^i_{pt}\boldsymbol{P}_{i-1},\boldsymbol{H}_{i-1}]\right), \quad i=1,2, \ldots, N.
\end{aligned}
\end{equation}

The gating function $g_{pt}$ is estimated from the Transformer input through a feedforward network with a sigmoid activation. The gate can precisely control the influence of the prompts for a specific Transformer layer. The impact of prompts will diminish if $g_{pt}$ gets close to 0.

Similarly, we design the gating function $\boldsymbol G_{A}$ to estimate the importance of the adapter for each layer. Considering that the proposed Inner-layer Adapter is inserted into a path parallel to the feedforward network, according to Eq. 4, the output of the Transformer layer consists of three components, that can be represented as: 

\begin{equation}
\begin{aligned}
{\boldsymbol H_{i}} &=\mathrm{LN}\left(\mathrm{FFN}\left(\boldsymbol{x_i}\right)+g^i_{A}(s \cdot \tilde{\boldsymbol{z}^{p}_i})+\boldsymbol{x_i}\right).
\end{aligned}
\end{equation}
 
The gating function $g^i_{A}$ is calculated from the hidden states $x_i$ via a FFN with sigmoid activation to estimate the importance of the adapter for the $i$-th layer. Thus, the adapter output is scaled by the gating function and fused with the other two branches. For the $i$-th layer, the Inner-layer Adapter is deactivated and bypassed if $g^i_{A}$ approaches 0. For Inter-layer Adapter, its output with the gating function is formulated as:                 

\begin{equation}
\begin{aligned}
\boldsymbol{\tilde{H}}=g^{N+1}_{A}(\mathrm{LN}(f(\boldsymbol{W}_{\text {inter}}(\sum_{i=1}^{N} w_{i} \boldsymbol H_{i}\label{con:DCT1D4})))),
\end{aligned}
\end{equation}
where the gating function $g^{N+1}_{A}$ is calculated from the hidden states after the weighted sum operation from all layers of the pre-trained model, via a FFN also with a sigmoid activation to regularize the adapter. 

Naively combining these two methods assigns each layer fixed PET methods, causing each of them to be applied equally. Alternatively, the UniPET-SPK has the flexibility to select and activate the appropriate PET methods based on tasks or datasets dynamically. In other words, the proposed framework has the ability to search and find a better mixture of the PET methods to achieve an overall enhanced performance.

\section{EXPERIMENTAL SETUP}
To investigate the performance of our proposed methods, we conduct a comprehensive series of experiments on three different datasets: VoxCeleb, CN-Celeb, and the significantly more challenging naturalistic field forensic 1$^{\rm{st}}$48-UTD corpus. In this section, we present dataset details, pre-trained speech models, speaker verification backends, model training details, and other compared methods.
\subsection{Dataset}
\subsubsection{\textbf{VoxCeleb}}
The VoxCeleb1\&2 datasets~\cite{nagrani2017voxceleb, chung2018voxceleb2} are collected from YouTube videos. For all experiments in Sec. VII-A, all ASV systems are trained on the VoxCeleb2 development set, which consists of 1,092,009 utterances from 5994 speakers. For all ablation study experiments in Sec. VII-B, all systems are trained on the VoxCeleb1 development set, which contains 148,642 utterances from 1211 speakers. We evaluate performance of all ASV systems on three evaluation trials: VoxCeleb1-O, VoxCeleb1-E, and VoxCeleb1-H. The VoxCeleb1-O is the original test trial list consisting of 37,611 trials from 40 speakers in VoxCeleb1 test set. The VoxCeleb1-E is the extended test trial list consisting of 579,818 trials from all 1,251 speakers in VoxCeleb1. The VoxCeleb1-H is the hard trial list consisting of 550,894 trials from 1,190 speakers in VoxCeleb1.    


\subsubsection{\textbf{CN-Celeb1}}
To investigate the robustness and generalization capabilities of the proposed PET methods, we also perform experiments on a Chinese dataset, CN-Celeb1. The CN-Celeb1~\cite{fan2020cn} dataset contains around 130,000 utterances from 1000 Chinese celebrities and covers 11 different genres, including entertainment, interviews, singing, play, movie, vlog, live broadcasts, speech, drama, recitation and advertisement. Most utterances include real-world noise: ambient noise, background babbling, music, cheers and laugh. We fine-tune the ASV systems on the CN-Celeb1 training set, which contains 111,260 utterances from 800 speakers, and evaluate on the evaluation set consisting of 18,224 utterances from 200 speakers. There are 3,604,800 pairs in the test trials. It is noted that CN-Celeb1 has a much smaller speaker/utterance size than VoxCeleb2. Additionally, domain mismatch exists between training and enrollment/test, which makes CN-Celeb1 a more challenging dataset for speaker verification. 

\subsubsection{\textbf{1$^{\textbf{st}}$48-UTD}}
In the final evaluation, the proposed PET methods are considered for the more challenging scenarios experienced in actual naturalistic forensics, based on the $1^{\rm{st}}$48-UTD corpus~\cite{sang2020open}. This dataset is derived from actual homicide investigations across U.S. cities documented in the TV program “The First 48”. \mfmod{The corpus is designed for forensic analysis of audio from various cities across the USA wherein detectives investigated actual homicides. The audio content was extracted from the USA TV program called “The First 48”, and all audio contents are recorded from various real locations (e.g., interview rooms, cars, fields). }This corpus contains 5,041 utterances from 300 speakers, totaling 3.5 hours of actual crime situational audio. \mfmod{In this dataset, each episode contains disjoint speakers, and speakers in each episode are tagged as Detective, Witness, and Suspect based on their identities. This is a small domain-specific dataset with short utterances. }The dataset includes utterances with an average length of 2.4 seconds, and over 50\% of these are shorter than 2 seconds. Besides short-duration utterances, context music, privacy based bleeps used for concealing some words, modified speech sounds, and some voice-overs are also included in the audio. After filtering the utterances consisting of non-speech content and speakers with fewer than three utterances, the training set used here comprises 3755 utterances from 228 speakers, while the test set includes 882 utterances from 39 speakers. We use the training portion of the $1^{\rm{st}}$48-UTD corpus for fine-tuning, and evaluate its performance on the test portion.

\subsection{Experimental Settings}
\noindent \textbf{Pre-trained Backbone}: In this study, we employ the WavLM Base+\footnote{https://huggingface.co/microsoft/wavlm-base-plus} and HuBERT Base\footnote{https://huggingface.co/facebook/hubert-base-ls960} as the pre-trained backbone models. They comprise a convolutional feature encoder and 12 Transformer blocks. Each Transformer block has 8 attention heads with 768-dimensional hidden states. The WavLM Base+ has a total of 94.70 million parameters and the HuBERT Base has a total of 94.68 million parameters. 

\noindent \textbf{SV Backend}: To demonstrate the potential for proposed PET methods to be robust and generalize well on different SV backends, we use two types of backends: (1) a simple backend consisting of two FC layers and a statistical pooling layer, with an embedding size of 512, and (2) a TDNN (x-vector)~\cite{snyder2018x} as the SV backend, similar to the settings in SUPERB~\cite{yang2021superb}. Here, we use the x-vector as the default backend if there is no specific explanation.  

\noindent \textbf{Inner+Inter Adapter}: The Inner-layer Adapter consists of two FC layers with a bottleneck dimension of 256 with ReLU activation function, followed by LN and a residual connection. The Inter-layer Adapter comprises one FC layer with a hidden dimension of 512, followed by a ReLU activation function and LN.  

\noindent \textbf{Deep Speaker Prompting}: To find the optimal speaker prompt length, we set a range of prompt lengths: $[1, 5, 10, 30, 50, 100]$. All the prompts are randomly initialized with Xavier uniform initialization scheme~\cite{glorot2010understanding}. In this study, we insert speaker prompts with the dimension of 768 to the input of every Transformer layer in the pre-trained model.

\noindent \textbf{UniPET-SPK}: We incorporate the Inner+Inter Adapter and Deep Speaker Prompting with the dynamic gating mechanism. Each Transformer layer is equipped with two distinct gating functions for Inner-layer Adapter and speaker prompts, respectively. Each gating function is computed by a separated FC layer with a hidden dimension of 768. We also assign one gating function for Inter-layer Adapter.      
 
All systems are trained with cross-entropy loss. We use the Adam~\cite{kingma2014adam} optimizer with an initial learning rate of 5e-4 for the SV backend and speaker prompt, and 1e-4 for all other parameters. We apply a warm-up strategy at the first 11.4k steps, with learning rates decreased to 1.5e-5 for SV backend and speaker prompt, and 3e-6 for all other parameters in the remaining steps.


\noindent \textbf{Comparing Methods}: We compare our proposed methods with the full fine-tuning and several transfer learning approaches, including:

\noindent \textbf{1)} \textbf{Conventional embedding extractor}: We choose x-vector~\cite{snyder2018x} and ECAPA-TDNN~\cite{desplanques2020ecapa} for comparison. We reimplement them and use training settings similar to those of other models. Besides, ECAPA-TDNN is trained with AAM-softmax loss~\cite{deng2019arcface}.

\noindent \textbf{2)} \textbf{Full fine-tuning}: We update all parameters of pre-trained SSL models but keep their convolutional encoder fixed.

\noindent \textbf{3)} \textbf{Backend}: This method only updates the SV backend while keeping the whole pre-trained model frozen.

\noindent \textbf{4)} \textbf{Weighted sum}: In addition to updating SV backend, this method also updates the learnable weights assigned for hidden states of all Transformer layers in the pre-trained model. We implement this following~\cite{yang2021superb}, which has shown this method to be effective for speaker verification. 

\noindent \textbf{5)} \textbf{Prameter-efficient tuning methods}: We consider (i) two adapter-based methods: Houlsby adapter~\cite{houlsby2019parameter} and E+L adapter~\cite{otake2023parameter} and (ii) Low-Rank Adaptation (LoRA)~\cite{hu2022lora} in this study. We use the same adapter dimension for these two adapter-based methods as our methods. For LoRA, we use the rank $r=64$, which achieves the best performance.

To ensure a fair comparison, we reimplement the Houlsby adapter and the E+L adapter, and use the same training configurations for these methods. We report the system performance using two evaluation metrics: Equal Error Rate (EER) and minimum Detection Cost Function (minDCF) with $p_{target}=0.05$.

\begin{table*}[t]
\centering
\caption{Performance of different transfer learning methods on VoxCeleb1. The second column represents the number of additional trainable parameters in the pre-trained model. Upper block: performance of all methods with the linear backend; Middle\&Lower block: performance of all methods with the TDNN backend; FT: full fine-tuning}
\setlength{\tabcolsep}{2.0mm}{
\renewcommand\arraystretch{1.2}
\scalebox{1.0}{
\begin{tabular}{lccccccccc}
\hline \multirow{2}{*}{\textbf{Method} } & \multirow{2}{*}{ \textbf{\# Params} } & \multicolumn{2}{c}{ \textbf{VoxCeleb1-O} } & & \multicolumn{2}{c}{ \textbf{VoxCeleb1-E} } & & \multicolumn{2}{c}{ \textbf{VoxCeleb1-H} } \\
\cline { 3 - 4 } \cline { 6 - 7 } \cline { 9 - 10 } & & \textbf{EER $(\%)$} & \textbf{minDCF} & & \textbf{EER(\%)} & \textbf{minDCF} & & \textbf{EER(\%)} & \textbf{minDCF} \\
\hline \hline 
x-vector & $6.1\mathrm{M}$ & 4.23 & 0.294 & & 4.56 & 0.295 & & 7.69 & 0.445 \\
ECAPA-TDNN & $14.7\mathrm{M}$ & 1.81 & 0.137 & & 1.92 & 0.139 & & 3.45 & 0.213 \\
\hline WavLM Base+ \& Linear Backend \\
\hline FT & $85.1\mathrm{M}$ $(90.0\%)$ + $0.2\mathrm{M}$ & 2.52 & 0.186 & & 2.65 & 0.185 & & 4.92 & 0.295 \\
Backend & $0.0 \mathrm{M}$ $(0.0 \%)$ + $0.2\mathrm{M}$ & 12.05 & 0.776 & & 14.21 & 0.872 & & 24.2 & 0.921 \\
Weighted Sum & $0.03\mathrm{M} $ $(0.03 \%)$ + $0.2\mathrm{M}$ & 4.01 & 0.316 & & 4.53 & 0.320 & & 8.03 & 0.470 \\  
LoRA {~\cite{hu2022lora}} & $2.4 \mathrm{M}$ $(2.5 \%)$ + $0.2\mathrm{M}$ & 4.14 & 0.329 & & 4.47 & 0.316 & & 7.75 & 0.424 \\
Houlsby Adapter {~\cite{houlsby2019parameter}} & $9.5 \mathrm{M}$ $(10.0 \%)$ + $0.2\mathrm{M}$ & 3.29 & 0.235 & & 3.91 & 0.242 & & 6.52 & 0.355 \\
E+L Adapter {~\cite{otake2023parameter}} & $9.1\mathrm{M}$ $(9.6 \%)$ + $0.2\mathrm{M}$ & 2.20 & 0.169 & & 2.59 & 0.176 & & 4.99 & 0.296 \\
\hline Inner-layer Adapter (ours) & $4.4\mathrm{M}$ $(4.6 \%)$ + $0.2\mathrm{M}$ & 2.39 & 0.181 & & 2.93 & 0.183 & & 5.09 & 0.298 \\
Inter-layer Adapter (ours) & $0.4\mathrm{M}$ $(0.4 \%)$ + $0.2\mathrm{M}$ & 3.58 & 0.271 & & 4.56 & 0.318 & & 8.79 & 0.473 \\
Inner+Inter Adapter (ours) & $4.8\mathrm{M}$ $(5.0 \%)$ + $0.2\mathrm{M}$ & $2.30$ & $0.179$ & & 2.62 & 0.179 & & 4.78 & 0.283 \\ 
Deep Speaker Prompting (ours) & $0.3 \mathrm{M}$ $(0.3 \%)$ + $0.2\mathrm{M}$ & $2.94$ & $0.218$ & & 2.75 & 0.184 & & 5.12 & 0.330 \\
UniPET-SPK (ours) & $5.1 \mathrm{M}$ $(5.4 \%)$ + $0.2\mathrm{M}$ & $\mathbf{2.11}$ & $\mathbf{0.162}$ & & $\mathbf{2.42}$ & $\mathbf{0.169}$ & & $\mathbf{4.58}$ & $\mathbf{0.271}$ \\
\hline WavLM Base+ \& TDNN Backend \\
\hline FT & $85.1\mathrm{M}$ $(90.0\%)$ + $5.1\mathrm{M}$ & 1.82 & 0.161 & & 2.08 & 0.142 & & 4.00 & 0.245 \\
Backend & $0.0 \mathrm{M}$ $(0.0 \%)$ + $5.1\mathrm{M}$ & 5.34 & 0.431 & & 5.57 & 0.343 & & 7.05 & 0.369 \\
Weighted Sum & $0.03 \mathrm{M} $ $(0.03 \%)$ + $5.1\mathrm{M}$ & 2.41 & 0.199 & & 2.71 & 0.179 & & 5.21 & 0.303 \\
LoRA {~\cite{hu2022lora}} & $2.4 \mathrm{M}$ $(2.5 \%)$ + $5.1\mathrm{M}$ & 2.35 & 0.179 & & 2.65 & 0.178 & & 5.04 & 0.288 \\
Houlsby Adapter {~\cite{houlsby2019parameter}} & $9.5 \mathrm{M}$ $(10.0 \%)$ + $5.1\mathrm{M}$ & 2.30 & 0.202 & & 2.77 & 0.204 & & 5.60 & 0.339 \\
E+L Adapter {~\cite{otake2023parameter}} & $9.1 \mathrm{M}$ $(9.6 \%)$ + $5.1\mathrm{M}$ & 1.86 & 0.151 & & 2.06 & 0.145 & & 4.26 & 0.264 \\
\hline Inner-layer Adapter (ours) & $4.4 \mathrm{M}$ $(4.6 \%)$ + $5.1\mathrm{M}$ & 1.94 & 0.165 & & 2.25 & 0.164 & & 4.34 & 0.264 \\
Inter-layer Adapter (ours) & $0.4 \mathrm{M}$ $(0.4 \%)$ + $4.7\mathrm{M}$ & 2.21 & 0.166 & & 2.55 & 0.173 & & 4.91 & 0.287 \\
Inner+Inter Adapter (ours) & $4.8 \mathrm{M}$ $(5.0 \%)$ + $4.7\mathrm{M}$ & $1.82$ & $0.156$  & & 2.18 & 0.150 & & 4.05 & 0.251 \\
Deep Speaker Prompting (ours) & $0.3 \mathrm{M}$ $(0.3 \%)$ + $5.1\mathrm{M}$ & $2.16$ & $0.172$  & & 2.41 & 0.169 & & 4.62 & 0.292 \\
UniPET-SPK (ours, w/o gate) & $5.1 \mathrm{M}$ $(5.4 \%)$ + $4.7\mathrm{M}$ & 1.80 & 0.148  & & 2.13 & 0.153 & & 4.11 & 0.252 \\
UniPET-SPK (ours) & $5.1 \mathrm{M}$ $(5.4 \%)$ + $4.7\mathrm{M}$ & $\mathbf{1.59}$ & $\mathbf{0.130}$  & & $\mathbf{1.74}$ & $\mathbf{0.138}$ & & $\mathbf{3.41}$ & $\mathbf{0.206}$ \\
\hline HuBERT Base \& TDNN Backend \\
\hline FT & $85.1\mathrm{M}$ $(90.0\%)$ + $5.1\mathrm{M}$ & 1.98 & 0.159 & & 2.12 & $\mathbf{0.147}$ & & 4.24 & 0.252 \\
Backend & $0.0 \mathrm{M}$ $(0.0 \%)$ + $5.1\mathrm{M}$ & 5.63 & 0.462 & & 5.82 & 0.355 & & 7.42 & 0.377 \\
Weighted Sum & $0.03 \mathrm{M} $ $(0.03 \%)$ + $5.1\mathrm{M}$ & 2.85 & 0.207 & & 3.02 & 0.210 & & 6.12 & 0.356 \\
LoRA {~\cite{hu2022lora}} & $2.4 \mathrm{M}$ $(2.5 \%)$ + $5.1\mathrm{M}$ & 2.55 & 0.187 & & 2.95 & 0.201 & & 5.94 & 0.345 \\
Houlsby Adapter {~\cite{houlsby2019parameter}} & $9.5 \mathrm{M}$ $(10.0 \%)$ + $5.1\mathrm{M}$ & 2.54 & 0.189 & & 2.96 & 0.209 & & 5.98 & 0.359 \\
E+L Adapter {~\cite{otake2023parameter}} & $9.1 \mathrm{M}$ $(9.6 \%)$ + $5.1\mathrm{M}$ & 2.25 & 0.176 & & 2.61 & 0.176 & & 5.10 & 0.301 \\
\hline Inner-layer Adapter (ours) & $4.4 \mathrm{M}$ $(4.6 \%)$ + $5.1\mathrm{M}$ & 2.13 & 0.165 & & 2.55 & 0.170 & & 5.07 & 0.300 \\
Inter-layer Adapter (ours) & $0.4 \mathrm{M}$ $(0.4 \%)$ + $4.7\mathrm{M}$ & 2.40 & 0.199 & & 2.79 & 0.202 & & 5.47 & 0.328 \\
Inner+Inter Adapter (ours) & $4.8 \mathrm{M}$ $(5.0 \%)$ + $4.7\mathrm{M}$ & $2.04$ & $0.159$  & & 2.37 & 0.163 & & 4.38 & 0.262 \\
Deep Speaker Prompting (ours) & $0.3 \mathrm{M}$ $(0.3 \%)$ + $5.1\mathrm{M}$ & $2.23$ & $0.175$  & & 2.72 & 0.200 & & 5.41 & 0.314 \\
UniPET-SPK (ours, w/o gate) & $5.1 \mathrm{M}$ $(5.4 \%)$ + $4.7\mathrm{M}$ & 2.01 & 0.161  & & 2.28 & 0.158 & & 4.40 & 0.265 \\
UniPET-SPK (ours) & $5.1 \mathrm{M}$ $(5.4 \%)$ + $4.7\mathrm{M}$ & $\mathbf{1.89}$ & $\mathbf{0.153}$  & & $\mathbf{2.06}$ & $0.148$ & & $\mathbf{4.10}$ & $\mathbf{0.246}$ \\
\hline
\end{tabular}}
}
\label{table:Ab:Vox2}
\end{table*}

\section{EXPERIMENTAL RESULTS}

\subsection{Comparison among Transfer Learning Methods Using English Data}
\mfmod{In these experiments, we investigate the performance of our proposed Inner+Inter Adapter, Deep Speaker Prompting, and our UniPET-SPK framework on VoxCeleb. The results presented in Table \ref{table:Ab:Vox2} consist of two parts: the upper portion shows results of all methods with a simple linear SV backend. The lower portion includes performance using the TDNN SV backend. For both SV backends, we observe a consistent performance trend, demonstrating that the proposed UniPET-SPK consistently outperforms full fine-tuning, Inner+Inter Adapter, Deep Speaker Prompting, and all other methods in all three VoxCeleb1 test trials. Compared to fine-tuning, UniPET-SPK improves performance with relative reductions of 10.7\% and 18.9\% for EER and minDCF respectively, while updating only 5.4\% of the parameters. Compared to Inner+Inter Adapter and Deep Speaker Prompting, the results show UniPET-SPK effectively integrates these methods with superior performance. This framework demonstrates compatibility between the Inner+Inter Adapter and Deep Speaker Prompting, allowing for dynamic optimization of method combinations for each layer. This mutually enhances performance and fully leverages speaker-related information embedded in all layers.}
In these experiments, we investigate the performance of our proposed Inner+Inter Adapter, Deep Speaker Prompting, and our UniPET-SPK framework on VoxCeleb. The results presented in Table \ref{table:Ab:Vox2} consist of three parts: the upper portion shows results of all methods with WavLM Base+ and a simple linear SV backend. The middle portion includes the performance with WavLM Base+ using the TDNN SV backend. The lower portion displays the performance of all methods with HuBERT Base followed by the TDNN SV backend. For all the different pre-trained SSL models and SV backends, we observe a consistent performance trend, demonstrating that the proposed UniPET-SPK consistently outperforms full fine-tuning, Inner+Inter Adapter, Deep Speaker Prompting, and all other methods in all three VoxCeleb1 test trials. Regarding the different SSL models, it is worth noting that UniPET-SPK achieves the best performance with WavLM Base+ using the TDNN backend. Compared to full fine-tuning, UniPET-SPK improves performance with relative reductions of 12.6\% and 19.3\% for EER and minDCF respectively, while updating only 5.4\% of the parameters. With HuBERT Base, UniPET-SPK consistently outperforms full fine-tuning with much fewer parameters. 

Compared to Inner+Inter Adapter and Deep Speaker Prompting, the results for both SSL models in Table I show that UniPET-SPK effectively integrates these methods with superior performance. The proposed framework demonstrates compatibility between the Inner+Inter Adapter and Deep Speaker Prompting, allowing for dynamic optimization of method combinations for each layer. Furthermore, the results of WavLM and HuBERT indicate that the gating mechanism significantly enhances performance compared to integrating adapters and prompts without gating. This highlights the crucial role of the gating mechanism in mutually improving submodule performance and fully leveraging speaker-related information embedded across all layers.

For the adapter-tuning method, our Inner+Inter Adapter consistently outperforms the Houlsby adapter and performs on par with E+L adapter with different SV backends for all three test trials. Moreover, we observe that both Inner-layer Adapter and Inter-layer Adapter with TDNN backend outperform the Houlsby adapter by a large margin with 53.7\% and 95.8\% fewer parameters. Compared to the weighted sum method, the Inter-layer Adapter only inserts one adapter after the weighted sum operation, achieving significantly better performance with the 8.3\% reduction in EER on VoxCeleb1-O. Although Deep Speaker Prompting slightly underperforms compared to full fine-tuning, it still exceeds several other PET methods including Houlsby and Inter-layer adapters, requiring only a minimal update of 0.3\% of the parameters. While LoRA has shown competitive performance when fine-tuning LLMs, there is still a large performance gap between LoRA and full fine-tuning for speaker verification. Experimental results demonstrate that the proposed adapter-tuning and prompt-tuning methods work well individually, and UniPET-SPK effectively combines both to achieve the best overall performance over the upper bound of either.  

\vspace{-2.0mm}
\subsection{Ablation Study}
In this section, we ablate the different design choices for Inner+Inter Adapter and Deep Speaker Prompting. All experiments for this ablation study are conducted on the VoxCeleb1 dataset.

\textbf{Adapter Dimension}. We conduct experiments to study how the bottleneck dimension of adapter impacts performance. As shown in Table \ref{table:Ab:dim}, Inner+Inter Adapter achieves the best performance when the dimension size is 256. We note that performance consistently improves with the bottleneck dimension increasing from 32 to 256, which indicates that an adapter with more parameters helps the model extract speaker information which is more discriminative. When the bottleneck dimension increases to 512, performance slightly degrades and reaches a saturation point, however this occurs with the number of parameters being almost doubled. Thus, we choose the more modest bottleneck dimension size of 256 in all experiments.     

\textbf{Adapter Scaling}. We further conduct experiments to study the effectiveness of the parallel adapter design. We compare the performance of our adapter framework using sequential and parallel insertion formulation. Results from Table \ref{table:Ab:scale} show the parallel adapter yields better performance than the sequential counterpart when using learnable and fixed scaling factors ($s \geq 0.5$). Moreover, we explicitly study the impact of the scaling factor on performance of the parallel adapter. Table \ref{table:Ab:scale} shows that the parallel adapter achieves the best performance with a fixed scale at 0.5. Using a learnable scale factor results in slightly inferior yet comparable results. Increasing or decreasing the value of $s$ causes a performance drop. The reason suggested is that a smaller $s$ might diminish the impact of task-specific features learned from the adapters, and a larger $s$ might weaken the contribution of task-agnostic features learned from the frozen pre-trained backbone. The experimental results demonstrate that the parallel adapter can be a better choice for speaker verification.

\textbf{Prompt token number}. To investigate the impact of prompt length on speaker verification, we conduct experiments to display how ASV performance changes with different lengths of speaker prompts. As shown in Table \ref{table:Ab:Prompt}, the Deep Speaker Prompting obtains the best performance when the length is 30. Increasing the prompt length from 1 to 30 brings performance improvement, while further increasing the length to 100 causes performance degradation. This suggests that a limited number of speaker prompts may not provide sufficient information and guidance to the pre-trained model. Also, increasing prompt length does not consistently give performance improvement. Therefore, we choose a prompt length of 30 in all experiments.       
  
\begin{table}[t]
\caption{Performance of Inner+Inter adapter with different hidden dimension sizes.}
\vspace{-4.0mm}
\setlength{\tabcolsep}{3mm}{
\begin{center}
\renewcommand\arraystretch{1.1}

\begin{tabular}{ll|cc}
\hline \makecell{\textbf{Dimension}} & {\textbf{\# Params}} & \textbf{EER} (\%) & \textbf{minDCF} \\
\hline
\makecell{FT} & \makecell{$85.1\mathrm{M}$} & $2.99$ & $0.228$ \\
\hline
\makecell{32} & \makecell{$0.9\mathrm{M}$} & $2.78$ & $0.187$ \\
\makecell{64} & \makecell{$1.5\mathrm{M}$} & $2.65$ & $0.184$ \\
\makecell{128} & \makecell{$2.6\mathrm{M}$} & $2.60$ & $0.180$ \\
\makecell{256} & \makecell{$4.7\mathrm{M}$} & $\mathbf{2.42}$ & $\mathbf{0.172}$ \\
\makecell{512} & \makecell{$9.1\mathrm{M}$} & $2.80$ & $0.201$ \\

\hline
\end{tabular}
\end{center}}
\label{table:Ab:dim}
\end{table}

\begin{table}[t]
\caption{Performance of sequential adapter and parallel adapter with alternate learnable and fixed scaling factors.}
\vspace{-4.0mm}
\setlength{\tabcolsep}{6mm}{
\begin{center}
\renewcommand\arraystretch{1.1}

\begin{tabular}{l|cc}
\hline \makecell{\textbf{Scales}} & \textbf{EER} (\%) & \textbf{minDCF} \\
\hline 
\makecell{FT} & $2.99$ & $0.228$ \\
\hline 
\makecell[c]{Sequential} & $2.94$ & $0.203$ \\ 
\makecell{Learnable} & $2.63$ & $0.181$ \\
\makecell{0.05} & $3.32$ & $0.221$ \\
\makecell{0.1} & $2.87$ & $0.205$ \\
\makecell{0.5} & $\mathbf{2.42}$ & $\mathbf{0.172}$ \\
\makecell{1.0} & $2.57$ & $0.180$ \\
\makecell{1.5} & $2.68$ & $0.191$ \\
\makecell{2.0} & $2.77$ & $0.198$ \\

\hline
\end{tabular}
\end{center}}
\label{table:Ab:scale}
\end{table}

\begin{table}[t]
\caption{Performance of deep speaker prompting with different prompt token lengths.}
\vspace{-2.0mm}
\setlength{\tabcolsep}{5mm}{
\begin{center}
\renewcommand\arraystretch{1.1}

\begin{tabular}{l|cc}
\hline \makecell{\textbf{Token Number}} & \textbf{EER} (\%) & \textbf{minDCF} \\
\hline
\makecell{FT} & $2.99$ & $0.228$ \\
\hline
\makecell{1} & $3.59$ & $0.269$ \\
\makecell{5} & $3.25$ & $0.243$ \\
\makecell{10} & $3.11$ & $0.232$ \\
\makecell{30} & $\mathbf{3.02}$ & $\mathbf{0.227}$ \\
\makecell{50} & $3.16$ & $0.234$ \\
\makecell{100} & $3.27$ & $0.239$ \\

\hline
\end{tabular}
\end{center}}
\label{table:Ab:Prompt}
\end{table}

\begin{figure}[t]
\centering
\scalebox{1.05}
{
\hspace{-6mm} \includegraphics[width=9.0cm,height=6.2cm]{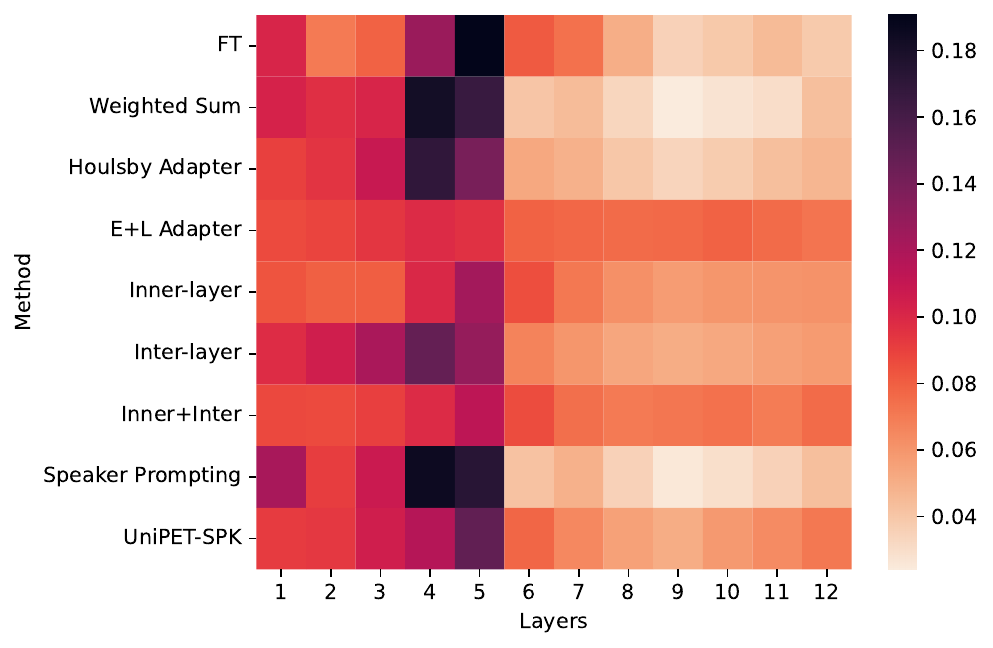}
}
\vspace{-9.0mm}
\caption{Layer weight analysis on the VoxCeleb dataset. Layers from 1 to 12 represent 12 Transformer layers in the WavLM Base+ model. Layer 1 is the one closest to the input.} 
\label{fig:Layer Weight}
\end{figure} 
\vspace{-4.0mm}

\subsection{Weight Analysis for Layer Contribution}
Following the settings used in SUPERB~\cite{yang2021superb}, the hidden states of all Transformer layers are weighted-sum with learnable weights and then fed to the SV backend. To understand how hidden states from different layers of a pre-trained model contribute to speaker verification across various transfer learning methods, we visualize the learnable weights for each Transformer layer obtained for these methods: full fine-tuning, weighted sum, Houlsby adapter, E+L adapter, Inner-layer Adapter, Inter-layer Adapter, Inner+Inter Adapter, Deep Speaker Prompting, and UniPET-SPK. As illustrated in Fig. \ref{fig:Layer Weight}, we observe a consistent pattern for all methods: the lower Transformer layers (e.g., Layer 1 to 6) contribute more to speaker verification, which agrees with findings from a separate study~\cite{chen2022wavlm}. \mfmod{Therefore, it may lose some significant speaker information if only fed the output of the last layer to the SV backend. }It is noted that if we neglect this info and use outputs from only the final layer, it could result in loss of crucial speaker information. In particular, we find the pattern that some bottom layers (Layer 2 and 3) and top layers (Layer 8 to 12) contribute slightly more in UniPET-SPK than full fine-tuning. This suggests that UniPET-SPK can better leverage more speaker-related information embedded in all layers, thereby it is beneficial to enhance speaker verification performance. The Inner+Inter adapter and Deep Speaker Prompting present different patterns. In the former, all layers contribute to the ASV task, while top layers contribute slightly less than bottom and middle layers. The reason could be that the interaction across all layers is considered in Inner+Inter adapter. Alternatively, top layers contribute much less than bottom layers in the Deep Speaker Prompting method. Compared to these two methods, UniPET-SPK learns characteristics from both and keeps the focus on Layer 3 to 5 while avoiding neglecting information from top layers. These tendencies therefore are factors that contribute to the superiority of UniPET-SPK. 

\mfmod{The Inner+Inter layer adapter and deep speaker prompting present a different pattern, which. These tendencies are the factors contributing to the superiority of UniPET-SPK. We also find the trend that the Inner+Inter Layer Adapter and deep speaker prompting also increase the middle layer's contribution. The analysis suggests that UniPET-SPK can better leverage more speaker-related information embedded in the hidden states of pre-trained models and improve speaker verification performance.} 

\subsection{Comparison among Transfer Learning Methods on Low-resource Chinese Dataset}
In this section, we demonstrate the performance of our approaches in a relatively low-resource scenario, using a challenging and smaller out-of-domain dataset. The WavLM model is pre-trained solely on English data. To present the robustness and generalization capabilities of our methods for non-English data, we conduct experiments and evaluate them on the CN-Celeb1 dataset. As shown in Table \ref{table:Ab:CNCeleb}, the UniPET-SPK achieves superior performance and outperforms full fine-tuning by 9.5\% and 7.0\% in EER and minDCF while only updating 5.4\% of the parameters. Moreover, UniPET-SPK consistently exceeds the performance of Inner+Inter Adapter and Deep Speaker Prompting. The results demonstrate the advantages of our UniPET-SPK framework over its submodules, and UniPET-SPK maintains excellent robustness and generalization ability across languages.  

Notably, on the CN-Celeb1 dataset, the Inner+Inter Adapter and Deep Speaker Prompting exhibit on par or slightly better performance than full fine-tuning. This suggests that our PET methods gain larger improvements and are more effective when training data is limited. Compared to other adapter-based methods, the Inner+Inter Adapter outperforms both Houlsby and E+L adapters with almost 2$\times$ fewer parameters. We also observe that the Deep Speaker Prompting performs slightly better than Houlsby and E+L adapters, while saving 96.8\% and 96.7\% of the parameters. These results indicate that UniPET-SPK offers clearly effective improvements for this challenging low-resource scenario compared to VoxCeleb.     
 
\begin{table}[t]
\centering
\caption{Performance of different transfer learning methods on CNCeleb1 Corpus.}
\setlength{\tabcolsep}{1.5mm}{
\renewcommand\arraystretch{1.4}
\scalebox{1.0}{
\begin{tabular}{lccc}
\hline \multirow{2}{*}{\textbf{Method} } & \multirow{2}{*}{ \textbf{\# Params} } & \multicolumn{2}{c}{ \textbf{CNCeleb1} } \\
\cline { 3 - 4 } & & \textbf{EER $(\%)$} & \textbf{minDCF} \\
\hline FT & $85.1\mathrm{M}$+$5.1\mathrm{M}$ & $14.49$ & $0.632$ \\
Backend & $0.0 \mathrm{M}$+$5.1\mathrm{M}$ & $20.12$ & $0.768$ \\
Weighted Sum & $0.03 \mathrm{M}$+$5.1\mathrm{M}$ & $15.90$ & $0.641$ \\
Houlsby Adapter {~\cite{houlsby2019parameter}} & $9.5 \mathrm{M}$+$5.1\mathrm{M}$ & $15.98$ & $0.683$ \\ 
E+L Adapter {~\cite{otake2023parameter}} & $9.1 \mathrm{M}$+$5.1\mathrm{M}$ & $15.46$ & $0.634$ \\
\hline Inner-layer Adapter (ours) & $4.4 \mathrm{M}$+$5.1\mathrm{M}$ & $14.52$ & $0.629$ \\
Inter-layer Adapter (ours) & $0.4 \mathrm{M}$+$4.7\mathrm{M}$ & $14.39$ & $0.630$ \\
Inner+Inter Adapter (ours) & $4.8 \mathrm{M}$+$4.7\mathrm{M}$ & $14.03$ & $0.612$ \\
Deep Speaker Prompting (ours) & $0.3 \mathrm{M}$+$5.1\mathrm{M}$ & $15.04$ & $0.631$ \\
UniPET-SPK (ours) & $5.1 \mathrm{M}$+$4.7\mathrm{M}$ & $\mathbf{13.12}$ & $\mathbf{0.588}$ \\
\hline
\end{tabular}}
}
\label{table:Ab:CNCeleb}
\end{table}

\begin{table}[t]
\caption{Performance of different transfer learning methods on $1^{st}$48-UTD forensic Corpus.}
\setlength{\tabcolsep}{1.5mm}{
\renewcommand\arraystretch{1.4}
\scalebox{1.0}{
\begin{tabular}{lccc}
\hline \multirow{2}{*}{\textbf{Method} } & \multirow{2}{*}{ \textbf{\# Params} } & \multicolumn{2}{c}{ \textbf{$1^{st}$48-UTD} } \\
\cline { 3 - 4 } & & \textbf{EER $(\%)$} & \textbf{minDCF} \\
\hline FT & $85.1 \mathrm{M}$+$4.5\mathrm{M}$ & $14.34$ & $0.671$ \\
Backend & $0.0 \mathrm{M}$+$4.5\mathrm{M}$ & $15.20$ & $0.689$ \\
Weighted Sum & $0.03 \mathrm{M}$+$4.5\mathrm{M}$ & $14.87$ & $0.680$ \\
Houlsby Adapter {~\cite{houlsby2019parameter}} & $9.5 \mathrm{M}$+$4.5\mathrm{M}$ & $13.71$ & $0.652$ \\ 
E+L Adapter {~\cite{otake2023parameter}} & $9.1 \mathrm{M}$+$4.5\mathrm{M}$ & $12.69$ & $0.613$ \\
\hline Inner-layer Adapter (ours) & $4.4 \mathrm{M}$+$4.5\mathrm{M}$ & $12.85$ & $0.621$ \\
Inter-layer Adapter (ours) & $0.4 \mathrm{M}$+$4.1\mathrm{M}$ & $13.01$ & $0.626$ \\
Inner+Inter Adapter (ours) & $4.8 \mathrm{M}$+$4.1\mathrm{M}$ & $12.32$ & $0.594$ \\
Deep Speaker Prompting (ours) & $0.3 \mathrm{M}$+$4.5\mathrm{M}$ & $13.25$ & $0.632$ \\
UniPET-SPK (ours) & $5.1 \mathrm{M}$+$4.1\mathrm{M}$ & $\mathbf{11.16}$ & $\mathbf{0.561}$ \\
\hline
\end{tabular}}
}
\label{table:Ab:1st48}
\end{table}

\subsection{Evaluation in More Challenging Naturalistic Forensic Scenarios}
A primary focus of this study has been to formulate advancement in robust speaker recognition specifically for forensic speaker applications. Therefore, to investigate the robustness and generalization capability of our methods, we address the more complex, challenging, and low-resource scenario of forensic speaker verification. The 1$^{st}$48-UTD forensic speaker recognition corpus is considered here. Table \ref{table:Ab:1st48} summarizes the performance of our proposed methods and other PET methods. The UniPET-SPK achieves the best performance and remarkably surpasses full fine-tuning and other methods. This demonstrates the advantages of our unified framework regarding model effectiveness and generalizability. Additionally, the Inner+Inter Adapter achieves the second-best performance, and the Deep Speaker Prompting performs significantly better than full fine-tuning. It is noted that limited data can lead to overfitting during fine-tuning. The results on 1$^{st}$48-UTD indicate that UniPET-SPK is quite robust and performs reliably under diverse scenarios. Improvements of UniPET-SPK over its submodules are generally larger when having fewer training samples, suggesting that UniPET-SPK performs especially well in the low-resource regime. \mfmod{The results illustrate effectiveness and robustness even in more complex scenarios, and parameter-efficient tuning methods perform even better in low-resource scenarios.}  

\section{Conclusion}
In this study, we explored how to efficiently and effectively fine-tune pre-trained self-supervised speech models for speaker verification. We proposed a parameter-efficient adapter-tuning method and a prompt-tuning method. To better incorporate these two methods, we also proposed UniPET-SPK, a unified parameter-efficient tuning framework that dynamically incorporates these two methods with a learnable gating mechanism to enhance the transfer of knowledge from pre-trained models toward the speaker verification task. \mfmod{aimed at effectively transferring the knowledge of pre-trained self-supervised speech models to the speaker verification task. }The UniPET-SPK framework was designed to achieve the optimal balanced combination of alternate PET methods and dynamically adjust the contribution of each method for individual layers. For the Inner+Inter Adapter, we proposed a parallel adapter design that incorporates two types of adapters to facilitate adaptation of both latent features within the intermediate Transformer layers and output embeddings from all Transformer layers. Our proposed Deep Speaker Prompting inserts learnable prompts into the latent space, thereby guiding the adaptation of the pre-trained model to speaker verification. Taking advantage of these two methods, the UniPET-SPK sufficiently leveraged the information embedded in all layers to find the optimal mixture of PET methods for each layer across a range of datasets and scenarios. 

We conducted comprehensive experiments on established VoxCeleb, CN-Celeb1, and the more challenging and naturalistic forensic speaker verification dataset $1^{\rm{st}}$48-UTD, to show that the UniPET-SPK can effectively incorporate Inner+Inter Adapter and Deep Speaker Prompting, and achieve better performance than full fine-tuning and other transfer learning methods. Our UniPET-SPK solution demonstrates its effectiveness and robustness for different pre-trained SSL models, languages, and complex scenarios. The proposed methods can efficiently adapt the pre-trained speech model to the downstream speaker verification task, leading to better performance with substantial reductions in computational and storage costs. It is suggested that this work could inspire future research on parameter-efficient transfer learning of large-scale pre-trained speech models for speaker verification. For future work, we may consider more large-scale pre-trained models for parameter-efficient fine-tuning methods on speech tasks.

\bibliographystyle{IEEEtran}
\bibliography{IEEEabrv,references}


\newpage

\vfill

\end{document}